\documentclass[twocolumn,showpacs,oneside,amsmath,amssymb,prl]{revtex4}
\usepackage{graphicx}
\usepackage{dcolumn}
\usepackage{bm}

\begin{document}

\title{Correlation-driven chiral superconductivity and chiral spin order in doped kagome lattice}

\author{Shun-Li Yu}
\affiliation{National Laboratory of Solid State
Microstructures and Department of Physics, Nanjing University,
Nanjing 210093, China}
\author{Jian-Xin Li}
\affiliation{National Laboratory of Solid State
Microstructures and Department of Physics, Nanjing University,
Nanjing 210093, China}

\date{\today}

\begin{abstract}
We study the electronic instabilities of the Hubbard model in the
1/6 hole-doped Kagome lattice using the variational cluster
approach. The 1/6 hole doping is unique in the sense that the
Fermi level is at the von Hove singularity and the Fermi surface
has a perfect nesting. In this case, a density wave is usually realized.
However, we demonstrate here that the chiral
$d_{x^{2}-y^{2}}+id_{xy}$  superconducting state is most favorable
when a small Hubbard interaction $U$($U<3.0t$) is introduced, and a scalar {\it chiral} spin order is
realized at large $U$($U>5.0t$). Between them, a spin-disordered
insulating state is proposed.

\end{abstract}


\maketitle

The kagome lattice has recently attracted considerable interest
due to its higher degree of frustration. Several possible states have
been proposed for the Heisenberg model in this lattice, including
the $U(1)$ algebraic spin liquid(SL)~\cite{ran}, the valance bond solid~\cite{Singh},
and the gapped SL~\cite{Jiang}. Recently, the numerical study shows that its ground state
is a singlet-gapped SL with signatures of $Z_{2}$ topological
order~\cite{Yan}. On the other hand, the anomalous quantum Hall effect~\cite{yu1}
and the topological
insulator~\cite{Guo} have also been demonstrated to exist when the
electron filling is near the Dirac point at 2/3 electron density.
In view of the rapid developments in the investigation of these exotic quantum phases,
the question arises if other correlation-driven exotic quantum orders will be
realized when the system is doped away from half-filling. We
note that the Fermi level is at the von Hove singularity and the
Fermi surface (FS) has a perfect nesting at 1/6 hole
doping. By using
the variational cluster approach(VCA) to the Hubbard model in the
kagome lattice, we show that a chiral
$d_{x^{2}-y^{2}}+id_{xy}$ superconductiving (SC) order and a
non-coplanar chiral spin density wave(SDW) can be realized
at the 1/6 hole doping.


Chiral superconductivity and chiral magnetic order are two
distinctive phases of matter. They break both the time-reversal
symmetry and the parity symmetry. The nontrivial topology of them
can result in a wealth of fascinating properties, such as
spontaneous quantum Hall effect~\cite{Horovitz,Ohgushi}, unusual
magnetoelectric properties~\cite{Bulaevskii}, and a quantized
boundary current in magnetic field~\cite{Laughlin}.
Experimentally, the spin-triplet $p$-wave chiral superconductivity
has been found in $\mathrm{Sr_{2}RuO_{4}}$ ~\cite{Mackenzie}. And,
the chiral spin order was proposed to describe the magnetic
ordering in $\mathrm{Mn}$ monolayers on $\mathrm{Cu(111)}$
surfaces~\cite{Kurz} and the nuclear spin ground state of a
two-dimensional solid $\mathrm{^{3}He}$~\cite{Momoi}. Recently,
the chiral spin order was also proposed in the doped triangular
lattice~\cite{Martin}, the pyrochlore lattice~\cite{Chern} and the
doped graphene~\cite{Li} based on the mean-field analysis. In
particular, the renormalization group
calculations~\cite{Chubukov,Wang} show that the chiral $d_{x^{2}-y^{2}}+id_{xy}$ superconductivity is favored in the
doped graphene with a perfect FS nesting. Here, we show
that both the $d_{x^{2}-y^{2}}+id_{xy}$ superconducting order and
a non-coplanar chiral SDW order can be realized in the 1/6 hole-doped kagome lattice by tuning the
on-site Coulomb interaction $U$.


The Hubbard model in the kagome lattice is defined as
\begin{eqnarray}
H=-t\sum_{\langle ij\rangle\sigma}(c^{\dag}_{i\sigma}c_{j\sigma}+h.c.)+U\sum_{i}n_{i\uparrow}n_{i\downarrow},
\end{eqnarray}
where $c^{\dag}_{i\sigma}$($c_{i\sigma}$) creates (annihilates) an
electron with spin $\sigma$ on site $i$ and
$n_{i\sigma}=c^{\dag}_{i\sigma}c_{i\sigma}$,
$\langle\cdot\cdot\rangle$ denotes the nearest-neighbor(NN) bond.
Each unit cell of the kagome lattice contains three sites (labeled
by $\alpha$, $\beta$ and $\gamma$ in Fig.1(a)). Except for a flat
band at the top of the energy band, the two dispersive bands are
the same as those in the honeycomb lattice (Fig.1(b)). From the
density of states shown in Fig.1(d), one can see that there are
three von Hove singularities at $-2t$, $0$ and $2t$. That at $2t$
originates from the flat band, the other two originate from the
saddle points at $M$ point (see Fig.1(b)). The Fermi levels at
$-2t$ and $0$ correspond to the $1/2$ and $1/6$ hole doping. In two cases, the FS forms a hexagon
and displays a perfect nesting(see Fig.1(c)). As the correlation
effect at $1/2$ doping is much weaker than that at $1/6$ doping,
we focus our study on the $1/6$ hole doping here.

VCA is a cluster approximation of the self-energy functional
approach~\cite{Potthoff}, which uses the rigorous variational
principle $\delta\Omega(\Sigma)/\delta\Sigma=0$ for the
thermodynamic grand potential $\Omega$ to determine the physical
self-energy $\Sigma$. It has been successfully applied to the
problem of competing phases in many strong correlation
systems~\cite{Senechal,Aichhorn,Sahebsara,Yu}. In VCA, the lattice
is tiled into identical clusters, which make up of a reference
system with the same two-body interaction as the original system
but a different one-body part(including the added Weiss fields to
study the symmetry broken phases). Then the exact Green function
$G^{\prime}$ (and the self-energy $\Sigma^{\prime}$) for each
cluster is calculated by exact diagonalization method. So, the
short-range static and dynamical (within each cluster)
correlations have been taken into account precisely. For any
$\Sigma^{\prime}$ parameterized as
$\Sigma^{\prime}(\mathbf{t}^{\prime})$, where
$\mathbf{t}^{\prime}$ represents the collection of the one-body
terms, we have the grand potential~\cite{Potthoff}:
\begin{eqnarray}
\Omega[\Sigma^{\prime}(\mathbf{t}^{\prime})]=\Omega^{\prime}(\mathbf{t}^{\prime})+\mathrm{Tr}
\ln[-G^{\prime}(\mathbf{t}^{\prime})]-\mathrm{Tr}\ln[-G(\mathbf{t}^{\prime})],
\label{grand potential}
\end{eqnarray}
where $\Omega^{\prime}(\mathbf{t}^{\prime})$ is the grand
potential of the reference system and $G(\mathbf{t}^{\prime})$ is
the approximate Green function of the original system calculated
through the cluster perturbation theory~\cite{Senechal1}.
Eq.(\ref{grand potential}) is no longer a functional but an
ordinary function of the variational parameters
$\mathbf{t}^{\prime}$, and the task of VCA is to find a stationary
point of this function,
$\partial\Omega(\mathbf{t}^{\prime})/\partial\mathbf{t}^{\prime}=0$.
In our calculation, the 12-site clusters(as enclosed by the dotted
lines in Fig.1(a)) and the open boundary conditions for the
clusters are used.
\begin{figure}
  \centering
  \includegraphics[scale=0.45]{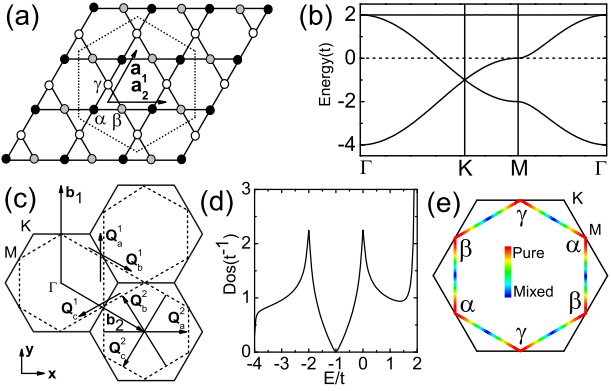}
  \caption{(color online) (a) Structure of kagome lattice and the 12-site cluster tiling (enclosed by doted lines) used in our VCA calculation.
  $\mathbf{a}_{1}$ and $\mathbf{a}_{2}$ are the lattice unit vectors. (b) The tight-binding dispersion along high symmetric directions(as illustrated in (c)). The dashed line is the Fermi
  level corresponding to the $1/6$ hole doping. (c) The Billouin zone and FS for the $1/6$ hole doping. $\mathbf{b}_{1}$ and $\mathbf{b}_{2}$ are the reciprocal-lattice vectors. The dashed lines denote the FS and the vectors $\mathbf{Q}^{1,2}_{a,b,c}$
  are the nesting vectors. (d) Density of state. (e) Weights of the contribution to
  FS from three inequivalent lattice sites $\alpha$, $\beta$ and $\gamma$ as represented by the colors. }
  \label{fig1}
\end{figure}

Before presenting our numerical results, let us first discuss the
possible SC symmetries at $1/6$ hole doping. We note that at each
saddle point $M$ the site-contribution to FS comes only from one
of the three inequivalent lattice sties, as shown in Fig1.(e).
Considering the effect of the van Hove singularity at $M$-point,
it is expected that the favorable Cooper pairings will be made
from two electrons belonging to the same sublattice. For the
kagome lattice, there are six coordinates in the same
sublattice for each site(Fig.1(a)), and it gives rise to three different bonds.
Considering that the paring intensities are the same along all
bonds, and the phase differences $\theta$ and $\phi$ with respect
to that along the ${\bf a}_{2}$ direction exists for the other two
directions, we will get the following pairing functions:
(\romannumeral1)
$\Delta(\mathbf{k})=\cos(k_{x})+e^{i\theta}\cos(k_{x}/2+\sqrt{3}k_{y}/2)+
e^{i\phi}\cos(k_{x}/2-\sqrt{3}k_{y}/2)$ for the spin-singlet
pairing, (\romannumeral2)
$\Delta(\mathbf{k})=\sin(k_{x})+e^{i\theta}\sin(k_{x}/2+\sqrt{3}k_{y}/2)+
e^{i\phi}\sin(k_{x}/2-\sqrt{3}k_{y}/2)$ for the spin-triplet
pairing. The most natural choices for the phases are
$(\theta,\phi)=(0,0)$ and $(2\pi/3,-2\pi/3)$. So, we have:
(\romannumeral1) the $d_{x^{2}-y^{2}}+id_{xy}$ wave for
$(\theta,\phi)=(2\pi/3,-2\pi/3)$,
$\Delta^{d+id}(\mathbf{k})=\Delta_{0}[\cos(k_{x})-\cos(k_{x}/2)\cos(\sqrt{3}k_{x}/2)+i\sqrt{3}\sin(k_{x}/2)\sin(\sqrt{3}k_{x}/2)]$;
(\romannumeral2) the extended $s$ wave for $(\theta,\phi)=(0,0)$,
$\Delta^{s}(\mathbf{k})=\Delta_{0}[\cos(k_{x})+2\cos(k_{x}/2)\cos(\sqrt{3}k_{x}/2)]$;
(\romannumeral3) the $p_{x}+ip_{y}$ wave for
$(\theta,\phi)=(2\pi/3,-2\pi/3)$,
$\Delta^{p+ip}(\mathbf{k})=\Delta_{0}[\sin(k_{x})-\sin(k_{x}/2)\cos(\sqrt{3}k_{x}/2)+i\sqrt{3}\cos(k_{x}/2)\sin(\sqrt{3}k_{x}/2)]$;
and (\romannumeral4) the $f$ wave for $(\theta,\phi)=(0,0)$,
$\Delta^{f}(\mathbf{k})=\Delta_{0}[\sin(k_{x})+2\sin(k_{x}/2)\cos(\sqrt{3}k_{x}/2)]$.
The phases and the gap nodes for the four pairing symmetries are
shown in Fig.2(a)-(d).
\begin{figure}
  \centering
  \includegraphics[scale=0.45]{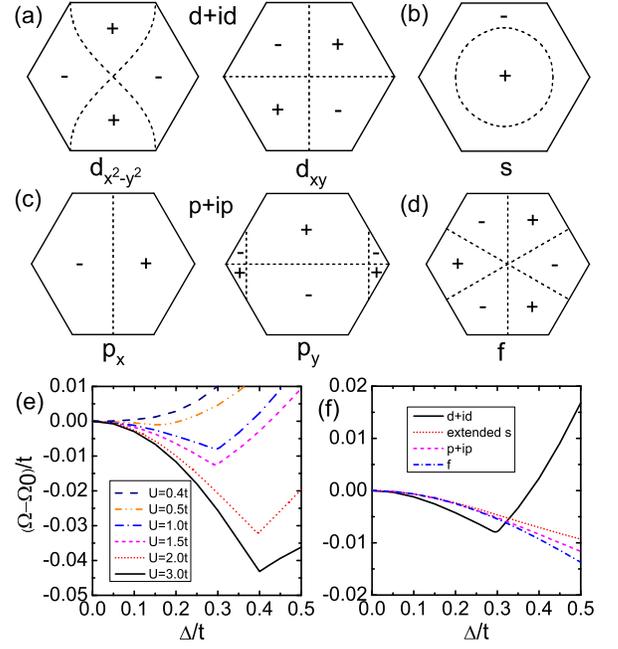}
  \caption{(color online) (a)-(d) Phase($\pm$) of the SC gap functions for the four possible pairing
  symmetries. The dotted lines denote the gap nodes.
  (e) Scaled grand potential $\Omega$ as a function of Weiss field $\Delta$ for different $U$
  in the $d_{x^{2}-y^{2}}+id_{xy}$ channel. (f) Scaled $\Omega$ as a function of $\Delta$ for
  different pairing symmetries at
  $U=t$.}
  \label{fig2}
\end{figure}

With the above pairing functions, we now introduce the Weiss
fields $H^{\prime}=\Delta\sum_{\langle\langle
ij\rangle\rangle}e^{i\theta_{ij}}\Delta^{s/t}_{ij}+h.c.$,
where $\Delta^{s/t}_{ij}=c_{i\uparrow}c_{j\downarrow}\mp c_{i\downarrow}c_{j\uparrow}$
for the singlet/triplet pairing, to test
the possible pairing orders. Here,
$\langle\langle\cdot\cdot\rangle\rangle$ denotes the bonds linking
the sites in the same sublattice and $\theta_{ij}$ is given
according to the rule discussed above. Fig.2(e) presents the
results of $\Omega-\Omega_{0}$ as a function of $\Delta$ for the
$d_{x^{2}-y^{2}}+id_{xy}$ pairing symmetry at various $U$.
$\Omega_{0}$ is the grand potential in the zero Weiss field. For
$0.4t<U<3t$, we find that there is a minimum at finite $\Delta$,
at the same time it satisfies
$\partial\Omega(\Delta)/\partial\Delta=0$. For $U\leq0.4t$, a
monotonic increase with $\Delta$ occurs. While, though there is
also a minimum at finite $\Delta$ for $U\geq3t$, we find that the
derivative at this minimum does not exist, because $\Omega$ is not
smooth at the point. Thus, it is not a stationary point of the
self-energy functional, according to the variational principle of
VCA~\cite{Potthoff}. So, the $d_{x^{2}-y^{2}}+id_{xy}$ pairing order
is realized only in the region $0.4t<U<3t$. We have
also checked the stationarity of other three SC pairing orders.
The typical results are shown in Fig.2(f) for $U=t$.
Except for the $d_{x^{2}-y^{2}}+id_{xy}$ pairing, all other three
pairings exhibit a monotonic decrease with the Weiss field. Therefore, they
are not stable solutions.


In the system with a perfect nesting FS, one will expect naturally an instability to
the density wave which may compete with the SC order. In the presence of the on-site Hubbard $U$, the most likely density wave would be a SDW.
A well known example is that the staggered SDW order occurs for the half-filled Hubbard model with a perfect FS nesting in the square lattice, where
the nesting wave vector $(\pm \pi, \pm \pi)$ coincides exactly the real-space translation symmetry of two unit distances for the staggered SDW.
Considering the geometry of the kagome lattice and consequently the nesting property for the 1/6 hole doping as shown in Fig.1(c), the staggered SDW
will not be favored here. Then, what is the kagome-lattice counterpart of the staggered SDW occurring on the half-filled square lattice?
At $1/6$ hole doping, the nesting FS has the nesting vectors $\mathbf{Q}^{2}_{a,b,c}$ (or $\mathbf{Q}^{1}_{a,b,c}$ which connects to $\mathbf{Q}^{2}_{a,b,c}$ via the reciprocal-lattice vectors) as indicated in Fig.1(c). With these nesting vectors, two distinct magnetic orders are expected. One is the non-coplanar chiral spin order associated with $\mathbf{Q}^{1}_{a,b,c}$ as shown in Fig.1(c), which is similar to that proposed in the doped triangle~\cite{Martin} and honeycomb~\cite{Li} lattices, where the four local spins direct along the normals to the faces of a regular tetrahedron(Fig.3). Thus, the spin orientations are:
$\mathbf{e}_{i\alpha}=1/\sqrt{3}[\mathbf{e}_{x}\cos(\mathbf{Q}^{1}_{a}\cdot\mathbf{R}_{i})+
\mathbf{e}_{y}\cos(\mathbf{Q}^{1}_{b}\cdot\mathbf{R}_{i})+\mathbf{e}_{z}\cos(\mathbf{Q}^{1}_{c}\cdot\mathbf{R}_{i})]$,
$\mathbf{e}_{i\beta}=1/\sqrt{3}[\mathbf{e}_{x}\cos(\mathbf{Q}^{1}_{a}\cdot\mathbf{R}_{i})-
\mathbf{e}_{y}\cos(\mathbf{Q}^{1}_{b}\cdot\mathbf{R}_{i})-\mathbf{e}_{z}\cos(\mathbf{Q}^{1}_{c}\cdot\mathbf{R}_{i})]$ and
$\mathbf{e}_{i\gamma}=1/\sqrt{3}[-\mathbf{e}_{x}\cos(\mathbf{Q}^{1}_{a}\cdot\mathbf{R}_{i})-
\mathbf{e}_{y}\cos(\mathbf{Q}^{1}_{b}\cdot\mathbf{R}_{i})+\mathbf{e}_{z}\cos(\mathbf{Q}^{1}_{c}\cdot\mathbf{R}_{i})]$,
where $\mathbf{R}_{i}$ denotes the position of the unit cell $i$. In this state, the resulting
scalar spin chirality, $\langle K_{s}\rangle=\langle\mathbf{S}_{\alpha}\cdot(\mathbf{S}_{\beta}\times\mathbf{S}_{\gamma})\rangle\ne 0$ in each triangular plaquette, breaks both the time-reversal and the parity symmetries. The other is the coplanar vector chiral spin order associated with $\mathbf{Q}^{2}_{a,b,c}$, as shown in Fig.3(b). In this case,
the spins in each triangular plaquette orient at $120^{\circ}$ to each other. The vector chirality for each triangle can be defined as
$\mathbf{K}_{v}=(2/3\sqrt{3})(\mathbf{S}_{\alpha}\times\mathbf{S}_{\beta}+\mathbf{S}_{\beta}\times\mathbf{S}_{\gamma}+
\mathbf{S}_{\gamma}\times\mathbf{S}_{\alpha})$, which is parallel to the $z$ axis
with amplitude $+1$ or $-1$. As shown in Fig.3(b), the vector chirality arranges as a \textquotedblleft stripe" phase, in which each chirality arranges in a stripe and the two stripes are staggered with each other. The scalar chiral spin order and the vector chiral spin order have different translational symmetries in real space as indicated in Fig.3(a) and (b), where the magnetic translation vectors are shown as the dotted lines.
\begin{figure}
  \centering
  \includegraphics[scale=0.45]{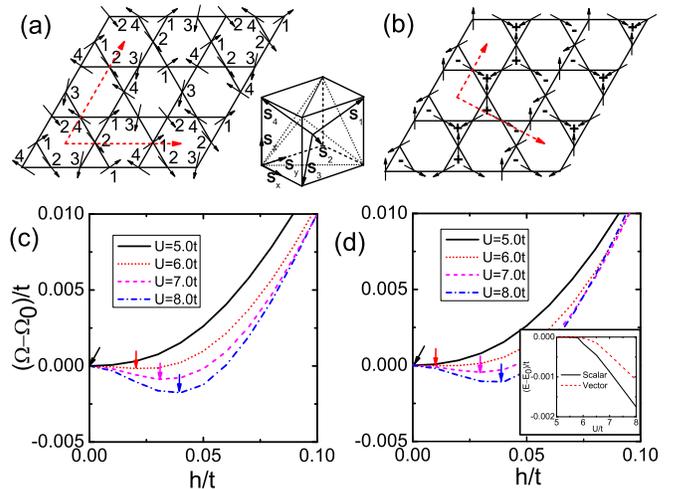}
  \caption{(color online) (a) Non-coplanar arrangement of spins(scalar spin chirality).
  $\mathbf{S}_{1\rightarrow4}$ indicate the four
  directions of local magnetic moments. (b) Coplanar arrangement of spins(vector spin chirality).
  The dashed lines with arrow in
  (a) and (b) are the magnetic translation vectors. (c) Scaled grand potential $\Omega$ as a function of
  the Weiss field $h$ for the spin order shown in (a) at various $U$. The local minima are indicated by arrows.
  (d) Scaled $\Omega$ as a function of
  $h$ for the spin order shown in (b). The inset of (d) shows the energy differences between the two spin ordered states
  and the normal phase with a zero Wiess field, respectively.}
  \label{fig3}
\end{figure}

To test the favorable SDW, we introduce the following Weiss field:
$H^{\prime}=h\sum_{i\eta}\mathbf{e}_{i\eta}\cdot\mathbf{S}_{i\eta}$, where $\mathbf{e}_{i\eta}$ is the orientation of magnetic moment and  $\mathbf{S}_{i\eta}=\sum_{\sigma\sigma^{\prime}}c^{\dag}_{i\eta\sigma}{\bm\tau}_{\sigma\sigma^{\prime}}c_{i\eta\sigma^{\prime}}$ with $\tau$ the Pauli matrixes, $i$ the unit cell index and $\eta$ the site index. Fig.3(c) and (d) show the results for $\Omega-\Omega_{0}$ as a function of the Weiss field $h$ for the scalar chiral spin order and the vector chiral spin order, respectively.
We find that a local minimum exists only for $U>5.5t$ in both cases. Therefore, in the region $0.4t<U<3t$, no SDW state will compete with the chiral $d_{x^{2}-y^{2}}+id_{xy}$ pairing state.
Fig.3(c) and (d) show that both spin ordering states are more favorable compared to the spin disordered state. So, we need to compare their energy gain which is given by the energy difference between the ordered solution and the normal solution found by suppressing the Weiss fields. The energy density $E=\Omega+\mu n$ as a function of electron density $n$ can be obtained for a solution with the functional $\Omega$.
The calculation of the energy gain is performed for several values of the chemical potential $\mu$ until the density $n$ is close enough to $1/6$ doping, and the cluster's chemical potential $\mu^{\prime}$ is also used as a variational parameter to guarantee the thermodynamic consistency~\cite{Aichhorn}. The inset of Fig.3(d) shows the energy gain for the two spin ordered solutions as a function of $U$.
It shows that the scalar chiral spin order state is more favorable than the vector chiral spin order state for $U>5.5t$. With decreasing $U$, the difference of the energy gain for the two spin-ordered states decreases.
Near the threshold $U$ for the occurrence of the spin order states, the two states are nearly degenerate.
\begin{figure}
  \centering
  \includegraphics[scale=0.45]{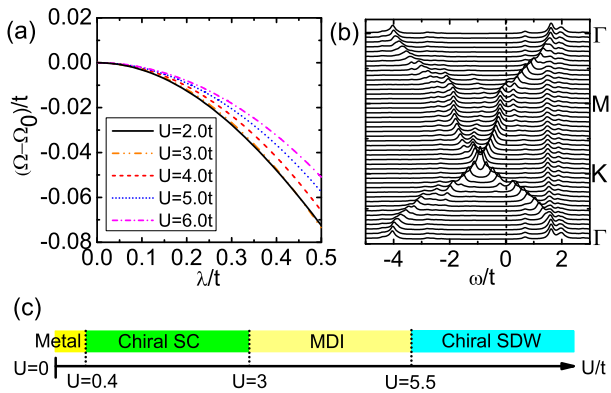}
  \caption{(color online) (a) Scaled grand potential $\Omega$ as a function of Weiss field $\lambda$ for the conventional CDW order (see text)
  at various $U$. (b) Spectral functions for $U=4t$ along the high symmetric directions(see Fig.1(c)). (c) Qualitative phase diagram of the Hubbard model in the kagome lattice for $1/6$ hole doping.}
  \label{fig4}
\end{figure}

To check the possible existence of the charge-density wave(CDW) in this system,
$H^{\prime}=\lambda\sum_{i}(f_{\alpha}n_{i\alpha}\cos(\mathbf{Q}^{1}_{a}\cdot\mathbf{r}_{i})+
f_{\beta}n_{i\beta}\cos(\mathbf{Q}^{1}_{b}\cdot\mathbf{r}_{i})+
f_{\gamma}n_{i\gamma}\cos(\mathbf{Q}^{1}_{c}\cdot\mathbf{r}_{i}))$ is used, with $i$ the unit cell index and $(f_{\alpha},f_{\beta},f_{\gamma})=(1,0,-1)$. The dependence of $\Omega-\Omega_{0}$ on the Weiss field $\lambda$ is presented in Fig4.(a). We find that $\Omega-\Omega_{0}$ decreases monotonously with $\lambda$ for
various $U$ from $U=2\sim 6t$. Therefore, the CDW does not exist in this region. Thus, in the region $0.4t<U<3t$, the only stable state we found is the chiral $d_{x^{2}-y^{2}}+id_{xy}$ pairing state, while for $U>5.5t$ the most favorable state is the scalar chiral spin order state. In the region between them, we do not find evidence of the SC state, spin order state and the CDW state. From the spectral function of single-particle excitations, we find that a gap occurs at the Fermi level as presented in Fig.4(b) for $U=4t$ (a non-vanishing spectral weight can be seen in the gapped region because a finite Lorentzian broadening is used in the
numerical calculations). So, the system is in fact an insulator. We would suggest that the system is a magnetic disordered insulator(MDI). The possible candidates may be the spin liquid, the spin glass or the valence-bond solid, but the method used here can not identify or distinguish them.

Finally, we summarize the results obtained above in the phase diagram shown in Fig4.(c). 
We note that the spin-$1/2$ kagome lattice has been realized in Herbertsmithite $\mathrm{Zn}\mathrm{Cu}_{3}\mathrm{(OH)}_{6}\mathrm{Cl}_{2}$~\cite{Shores,Lee} and its isostructural $\mathrm{Mg}$-based paracatamite $\mathrm{Mg}\mathrm{Cu}_{3}\mathrm{(OH)}_{6}\mathrm{Cl}_{2}$~\cite{Kermarrec}. Also, the kagome lattice has been simulated experimentally in ultra-cold atoms~\cite{Jo}. So,
we suggest that the theoretical predictions presented here may be probed after doping these compounds or by implementing an optical lattice in ultra-cold atoms where the Hubbard interaction $U$ can be tuned continuously. 

In summary, we have studied the Hubbard model in the $1/6$ hole doped kagome lattice using the variational cluster approach.
We find that the chiral $d_{x^{2}-y^{2}}+id_{xy}$ superconducting state is most favorable at a small Hubbard interaction $U$($0.4t<U<3t$),
and the scalar chiral spin order is realized at the large $U$($U>5.5t$). Between them, a magnetic disordered insulating state is proposed.

\begin{acknowledgments}
We thank F. Wang and Q. H. Wang for helpful discussions, in particular F. Wang
for pointing out Ref.\cite{Chubukov}. This work was supported by the National Natural Science Foundation of
China (10525415) and the Ministry of Science and Technology of
China (973 project Grants Nos.2006CB601002,2006CB921800).
\end{acknowledgments}

\end{document}